\documentclass[10pt,preprint2]{aastex}

\usepackage{graphicx}
\def \apj {ApJ}
\def \apjs {ApJS}
\def \apjl {ApJ}

\def \aap {A\&A}

\newcommand{\citeN}[1]{\citeauthor{#1} (\citeyear{#1})}
\newcommand{\citeNP}[1]{\citeauthor{#1} \citeyear{#1}}

\shortauthors{Socas-Navarro \& Trujillo Bueno}
\shorttitle{Polynomial Approximants for \ion{He}{1} 10830}

\begin{document}

\title{Polynomial Approximants for the Calculation of Polarization Profiles
  in the \ion{He}{1}  10830~\AA \, Multiplet}
%Zeeman Components and Strengths of the \ion{He}{1} 10830 \AA \, Multiplet in the Paschen-Back Effect Regime}

\author{H. Socas-Navarro}
   	\affil{High Altitude Observatory, NCAR\thanks{The National Center
	for Atmospheric Research (NCAR) is sponsored by the National Science
	Foundation.}, 3450 Mitchell Lane, Boulder, CO 80307-3000, USA}
	\email{navarro@ucar.edu}

\author{J. Trujillo Bueno	\thanks{Consejo Superior de Investigaciones
    Cient\' \i ficas, Spain} 
} 
   	\affil{Instituto de Astrof\'\i sica de Canarias, Avda V\'\i a L\'
   	actea S/N, La Laguna 38205, Tenerife, Spain}
	\email{jtb@iac.es}

\author{E. Landi Degl'Innocenti}
        \affil{Dipartimento di Astronomia e Scienza dello Spazio, Largo
        E. Fermi 2, 50125 Firenze, Italy}
	\email{landie@arcetri.astro.it}

\date{}%

\begin{abstract}
%
% HSN
The \ion{He}{1} multiplet at 10830 \AA \, is formed in the incomplete
Paschen-Back regime for typical conditions found in solar and stellar
atmospheres.  The positions and strengths of the various components that form
the Zeeman structure of this multiplet in the Paschen-Back regime are
approximated here by polynomials. The fitting
%The positions and strengths of the various components that form the Zeeman
%structure of the \ion{He}{1} multiplet at 10830 \AA \, in the incomplete
%Paschen-Back regime are approximated by third-order polynomials. The fitting
errors are smaller than $\sim$10$^{-2}$ m\AA \, in the component positions
and $\sim$10$^{-3}$ in the relative strengths. The approximant polynomials
allow for a very fast implementation of the incomplete Paschen-Back regime in
numerical codes for the synthesis and inversion of polarization profiles in
this important multiplet.
\end{abstract}
   

\section{Introduction}
\label{intro}

% HSN
% spelled out IR
One of the most useful near-infrared spectral regions for spectropolarimetric
investigations of solar and stellar magnetic fields is that located around
10830 \AA, which contains both the photospheric line of \ion{Si}{1} at 10827
\AA\ and the chromospheric lines that result from the \ion{He}{1} 10830 \AA\
multiplet (\citeNP{HH71}; \citeNP{RSL95}; % \citeNP{LPK98};
\citeNP{TBLdIC+02}). Of particular interest is the fact that the 
`blue' and `red' components of this multiplet are sensitive to both the Hanle
and Zeeman effects, which offers a suitable diagnostic window for
%
% HSN
% Removed 1 ``solar'' 
investigating magnetic fields in a variety of solar plasma structures,
such as prominences (\citeNP{TBLdIC+02}), emerging flux regions
(\citeNP{SLW+03}) and solar chromospheric spicules (\citeNP{TBMC+05};
\citeNP{SNE05}). 
%We recall that t
%
% HSN
The \ion{He}{1} 
10830~\AA \ multiplet originates between a lower term ($2^3{\rm S}_1$) and an
upper term ($2^3{\rm P}_{2,1,0}$).  Therefore, it comprises three spectral
lines (see, e.g., \citeNP{RS85}): a `blue' component at 10829.09~\AA\ with
$J_l=1$ and $J_u=0$ (hereafter Transition~1, or Tr1 for abbreviation), and
two `red' components at 10830.25~\AA\ with $J_u=1$ (hereafter, Tr2) and at
10830.34~\AA\ with $J_u=2$ (hereafter, Tr3) which appear blended at solar
atmospheric temperatures.

In a previous paper (\citeNP{SNTBLdI04}, hereafter paper-I), we demonstrated
that the determination of the magnetic field vector via the analysis of the
observed polarization in the \ion{He}{1} 10830~\AA \,
%
% HSN
multiplet must be carried out considering the wavelength positions and the
strengths of the Zeeman components in the incomplete Paschen-Back effect
regime. 
The theory of the Paschen-Back effect in this regime is
detailed in Section 3.4 of the book by \citeN{LdIL04}. The problem of finding
the strengths and the splittings of the various magnetic components, arising
in the transition from a lower and an upper term, both described in the L-S
coupling scheme ($^3$S and $^3$P, respectively, in the case of the
$\lambda$10830 line), is reduced to the numerical diagonalization of a number
of matrices. The expressions of the matrix elements are given, in terms of
Wigner-symbols, in Eq.~(3.57) of that book, or, in analytical
form, in Eqs.~(3.61a,b). As pointed out in paper-I,
our results have been obtained through the adaptation to the
fine-structure case of a computer program previously developed to handle the
similar problem for hyperfine-structured multiplets (\citeNP{LdI78}).

\begin{figure*}
\epsscale{1.5}
\plotone{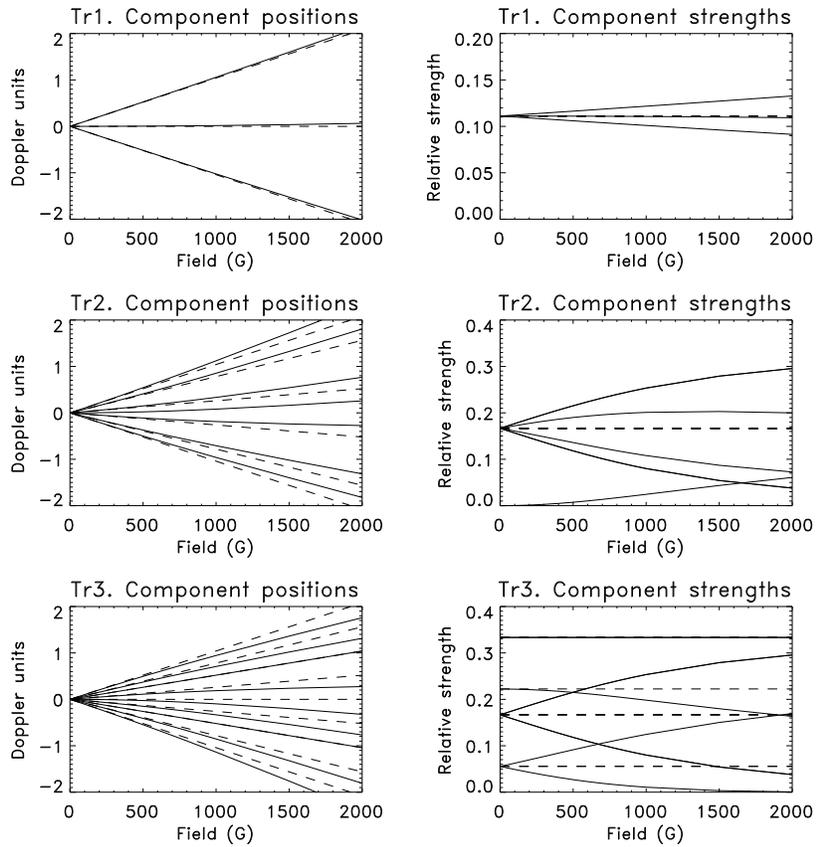}
\caption{Positions and strenghts of the Zeeman components
  as a function of the magnetic field. In all panels the solid lines
  represent calculations considering IPBS while the dashed lines represent
  those using the LZS approximation. Note that, in the general IPBS case, the
  Zeeman components exhibit asymmetric displacements and strengths. Also notice
  that the Zeeman IPBS components have strengths that depend on
  the magnetic field.
The central $\pi$-component of Tr2 has
zero strength in LZS and is not plotted in the figure. The ordinate axes are
given in Doppler units (corresponding to 105~m\AA ) to make the figure
directly comparable with that in paper-I. 
\label{fig:corrected}
}
\end{figure*}

Fig~\ref{fig:corrected} shows the transition components with the linear
Zeeman splitting (LZS) approximation and the rigorous calculation considering
the incomplete Paschen-Back splitting effects (IPBS). The figure has been
reproduced from paper-I, but corrects a plotting error in one of the Tr3
$\pi$-components. 

%, as done by \citeN{TBLdIC+02} in their investigation of the
%magnetic field that confines the plasma of solar coronal filaments. 
%
% HSN
%Actually,

Significant differences between LZS and IPBS are to be expected for typical
magnetic strengths in 
sunspots, simply because for fields stronger than about 400 gauss the Zeeman
splittings of the $J=2$ and $J=1$ levels of the upper term become
% 
% HSN
comparable to their energy separation and the perturbation theory of the
familiar Zeeman effect is no longer valid. However, the discrepancies between
the emergent Stokes profiles calculated assuming LZS or IPBS turn out to be
significant 
even for field strengths substantially weaker than the level-crossing field
of $\sim$400~G (see paper-I).

In this paper we provide polynomial approximants that allow researchers to
calculate Zeeman components and strengths of this interesting multiplet in
the IPBS regime.  This will facilitate the generalization of numerical codes
developed with the LZS approximation, such as the Milne-Eddington code of
\citeNP{LWK+04}, and to improve the efficiency of the IPBS code of
Socas-Navarro et al (2004). Moreover, it may also be useful for solving the
full non-LTE polarization transfer problem in the \ion{He}{1} 10830 \AA\
multiplet within the framework of the incomplete Paschen-Back effect theory.

\begin{figure*}
\epsscale{1.5}
\plotone{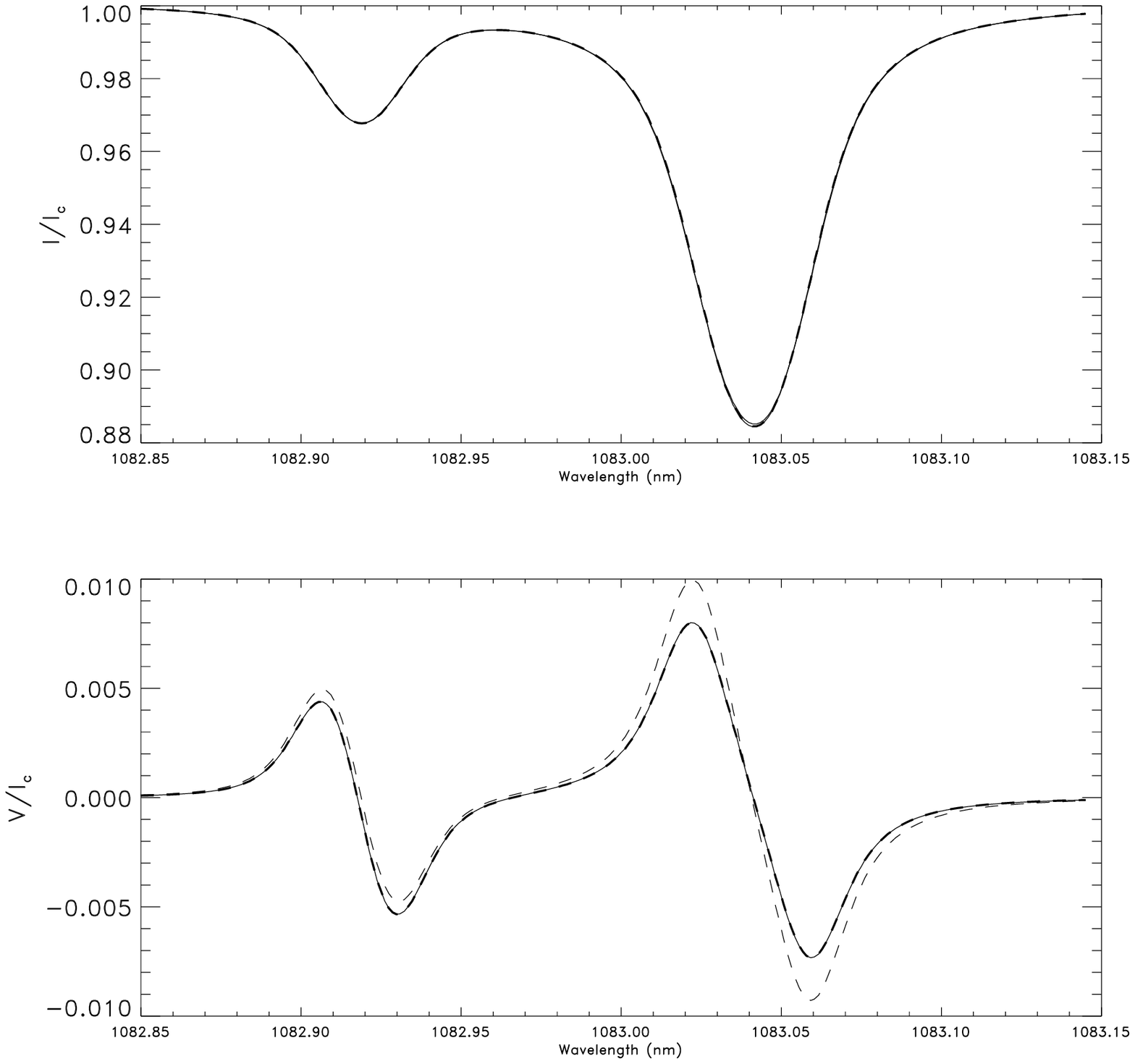}
\caption{
Stokes~I and~V profiles calculated according to a Milne-Eddington atmosphere
for a 500~G field. Solid line: 
Strict   IPBS calculation. Thin dashed line: LZS approximation. Thick dashed
line: LZS corrected with the coefficients provided in this paper. The
error induced by this approximation on the synthetic profiles is smaller
than $\sim$10$^{-4}$ and is hardly noticeable in the figure. 
\label{f1}
}
\end{figure*}

%The aim of the present 
%paper is to provide such polynomial approximants for the 
%fast calculation of Zeeman components and strengths in the Paschen-Back
%regime. 
Figure~\ref{f1} demonstrates that our polynomial approach
to the IPBS regime in the \ion{He}{1} 10830 \AA\ multiplet is
sufficiently accurate for the calculation of the emergent Stokes profiles
from a magnetized stellar atmosphere. 

\section{Polynomial fits}
\label{pol}

Let us characterize the deviation of the Zeeman pattern from the LZS
as polynomials. For each Zeeman component, we have:
\begin{eqnarray}
\Delta \lambda_{IPB} = \Delta \lambda_{LZS} + y_p(x) \nonumber \\
f_{IPB}=f_{LZS} + y_f(x) \, ,
\end{eqnarray}
where $\Delta \lambda$ denotes the relative position of a given component (in
m\AA ), $f$ represents the relative strengths which are dimensionless
quantities (see, e.g., Eq~[3.64] in \citeNP{LdIL04}), $x$ is the magnetic field
intensity expressed in kG, and $y(x)$ are the approximant polynomials. Since
the expressions for the splittings and the strengths in the IPBS regime have
to converge to those of the LZS effect 
under the limit of weak fields, and since, under the same limit, the
splittings are linear in the magnetic field and the strengths tend to a
constant, it has to be expected that the differences $y_p$ and $y_f$ are of
the form:
\begin{eqnarray}
y_p= c_2 x^2 + c_3 x^3 + c_4 x^4 + c_5 x^5 \,  \nonumber \\
y_f= d_1 x   + d_2 x^2 + d_3 x^3 + d_4 x^4 \, ,
\end{eqnarray}
with $0 < x < 3$. 
%The strenghts $f$ are dimensionless parameters that
%account for the different relative transition probabilities of the
%component (see, e.g. Eq~[3.64] of \cite{LdIL04}).

% The absence of an independent term in the polynomial
%ensures that one recovers the LZS regime (i.e., $y \rightarrow 0$) for $x
%\rightarrow 0$ (\citeNP{LdIL04}).

\begin{figure*}
\epsscale{1.5}
\plotone{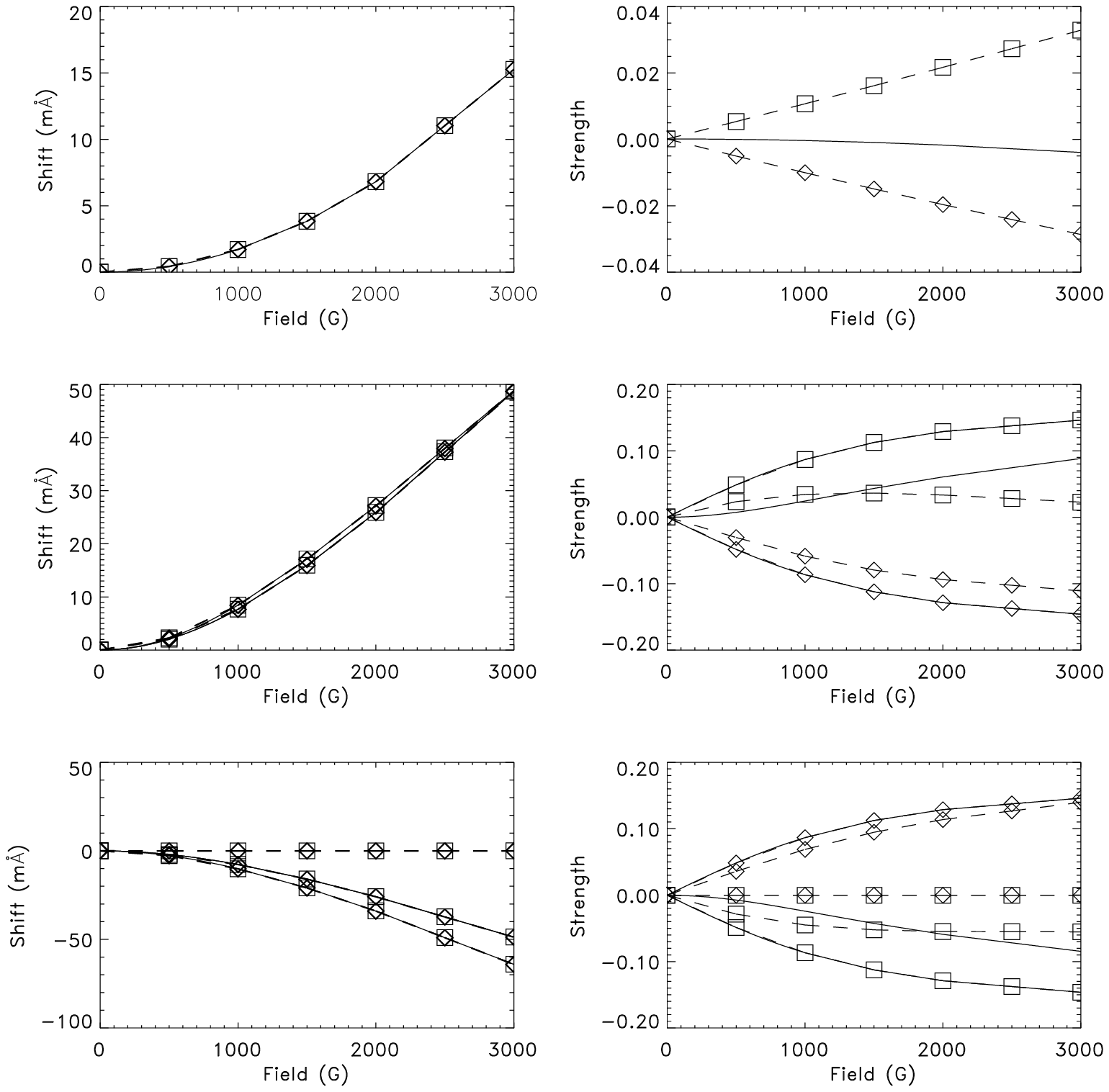}
\caption{Difference between IPBS and LZS component. Left panels: Position
  shifts ($\Delta \lambda_{IPB} - \Delta \lambda_{LZS}$). Right panels:
  Relative strengths ($f_{IPB} - f_{LZS}$). Top to bottom: Tr1, Tr2 and Tr3. 
In all panels, the solid lines represent $\pi$-components and the
  dashed lines represent $\sigma$-components (diamonds: $\sigma^+$, squares:
  $\sigma^-$). 
\label{f2}
}
\end{figure*}

The following tables list the coefficients for the corrections $y_p(x)$ and
$y_f(x)$ that need to be applied to the LZS approximation to account
for IPB effects. 
The last column of each table lists the maximum fitting error in the range
considered ($0<x<3$).
The components are sorted in the order of increasing
wavelengths in the LZS regime. 
% (see \citeNP{LdIL04} for a detailed
%presentation of the Zeeman and Paschen-Back theories). 

%%%%%%%%%%%%%%%%%%%%%%%%%%

\begin{deluxetable}{lccccc}
\tablewidth{0pt}
\tablecaption{Component positions. Tr1}
\label{pos_Tr1}
\tablehead{Transition & $c_2$ & $c_3$ & $c_4$ & $c_5$ & Max error  }
\startdata
$\sigma^+$ &    1.685  &    3.072$ \times 10^{-2} $  &   -1.366$ \times 10^{-2} $  &    1.568$ \times 10^{-3} $  &    4.508$ \times 10^{-3} $  \\
$\pi$ &    1.685  &    3.073$ \times 10^{-2} $  &   -1.366$ \times 10^{-2} $  &    1.567$ \times 10^{-3} $  &    4.505$ \times 10^{-3} $  \\
$\sigma^-$ &    1.685  &    3.073$ \times 10^{-2} $  &   -1.366$ \times 10^{-2} $  &    1.567$ \times 10^{-3} $  &    4.505$ \times 10^{-3} $  \\
\enddata
\end{deluxetable}

%%%%%%%%%%%%%%%%%%%%%%%%%%

\begin{deluxetable}{lccccc}
\tablewidth{0pt}
\tablecaption{Component strengths. Tr1}
\label{str_Tr1}
\tablehead{Transition & $d_1$ & $d_2$ & $d_3$ & $d_4$ & Max error }
\startdata
$\sigma^+$ &   -1.040$ \times 10^{-2} $  &    2.326$ \times 10^{-4} $  &    1.577$ \times 10^{-5} $  &   -8.117$ \times 10^{-7} $  &    6.836$ \times 10^{-7} $  \\
$\pi$ &   -1.280$ \times 10^{-7} $  &   -4.635$ \times 10^{-4} $  &    2.127$ \times 10^{-8} $  &    1.606$ \times 10^{-6} $  &    1.034$ \times 10^{-7} $  \\
$\sigma^-$ &    1.040$ \times 10^{-2} $  &    2.331$ \times 10^{-4} $  &   -1.740$ \times 10^{-5} $  &   -4.781$ \times 10^{-7} $  &    8.643$ \times 10^{-7} $  \\
\enddata
\end{deluxetable}

%%%%%%%%%%%%%%%%%%%%%%%%%%

\begin{deluxetable}{lccccc}
\tablewidth{0pt}
\tablecaption{Component positions. Tr2}
\label{pos_Tr2}
\tablehead{Transition & $c_2$ & $c_3$ & $c_4$ & $c_5$ & Max error  }
\startdata
$\sigma^+$ &    8.477  &   -4.631$ \times 10^{-1} $  &   -4.402$ \times 10^{-1} $  &    8.436$ \times 10^{-2} $  &    9.841$ \times 10^{-3} $  \\
$\sigma^+$ &    9.678  &   -8.777$ \times 10^{-1} $  &   -4.968$ \times 10^{-1} $  &    1.052$ \times 10^{-1} $  &    1.569$ \times 10^{-2} $  \\
$\pi$ &    8.556  &   -5.492$ \times 10^{-1} $  &   -4.028$ \times 10^{-1} $  &    7.882$ \times 10^{-2} $  &    8.270$ \times 10^{-3} $  \\
$\pi$ &    9.757  &   -9.627$ \times 10^{-1} $  &   -4.601$ \times 10^{-1} $  &    9.984$ \times 10^{-2} $  &    1.419$ \times 10^{-2} $  \\
$\pi$ &    8.517  &   -5.067$ \times 10^{-1} $  &   -4.212$ \times 10^{-1} $  &    8.153$ \times 10^{-2} $  &    9.087$ \times 10^{-3} $  \\
$\sigma^-$ &    9.835  &   -1.049  &   -4.226$ \times 10^{-1} $  &    9.430$ \times 10^{-2} $  &    1.264$ \times 10^{-2} $  \\
$\sigma^-$ &    8.596  &   -5.927$ \times 10^{-1} $  &   -3.839$ \times 10^{-1} $  &    7.602$ \times 10^{-2} $  &    7.527$ \times 10^{-3} $  \\
\enddata
\end{deluxetable}

%%%%%%%%%%%%%%%%%%%%%%%%%%

\begin{deluxetable}{lccccc}
\tablewidth{0pt}
\tablecaption{Component strengths. Tr2}
\label{str_Tr2}
\tablehead{Transition & $d_1$ & $d_2$ & $d_3$ & $d_4$ & Max error }
\startdata
$\sigma^+$ &   -1.019$ \times 10^{-1} $  &    4.219$ \times 10^{-3} $  &    1.334$ \times 10^{-2} $  &   -2.952$ \times 10^{-3} $  &    1.464$ \times 10^{-3} $  \\
$\sigma^+$ &   -5.746$ \times 10^{-2} $  &   -1.458$ \times 10^{-2} $  &    1.583$ \times 10^{-2} $  &   -2.904$ \times 10^{-3} $  &    1.014$ \times 10^{-3} $  \\
$\pi$ &    1.019$ \times 10^{-1} $  &   -4.220$ \times 10^{-3} $  &   -1.334$ \times 10^{-2} $  &    2.952$ \times 10^{-3} $  &    1.464$ \times 10^{-3} $  \\
$\pi$ &   -2.566$ \times 10^{-4} $  &    3.714$ \times 10^{-2} $  &   -1.443$ \times 10^{-2} $  &    1.785$ \times 10^{-3} $  &    4.085$ \times 10^{-4} $  \\
$\pi$ &   -1.019$ \times 10^{-1} $  &    4.219$ \times 10^{-3} $  &    1.334$ \times 10^{-2} $  &   -2.952$ \times 10^{-3} $  &    1.464$ \times 10^{-3} $  \\
$\sigma^-$ &    5.772$ \times 10^{-2} $  &   -2.256$ \times 10^{-2} $  &   -1.398$ \times 10^{-3} $  &    1.118$ \times 10^{-3} $  &    1.422$ \times 10^{-3} $  \\
$\sigma^-$ &    1.019$ \times 10^{-1} $  &   -4.220$ \times 10^{-3} $  &   -1.334$ \times 10^{-2} $  &    2.952$ \times 10^{-3} $  &    1.464$ \times 10^{-3} $  \\
\enddata
\end{deluxetable}

%%%%%%%%%%%%%%%%%%%%%%%%%%

\begin{deluxetable}{lccccc}
\tablewidth{0pt}
\tablecaption{Component positions. Tr3}
\label{pos_Tr3}
\tablehead{Transition & $c_2$ & $c_3$ & $c_4$ & $c_5$ & Max error  }
\startdata
$\sigma^+$ &   -1.603$ \times 10^{-2} $  &    4.348$ \times 10^{-3} $  &    6.113$ \times 10^{-4} $  &   -2.495$ \times 10^{-4} $  &    3.138$ \times 10^{-3} $  \\
$\sigma^+$ &   -8.602  &    5.997$ \times 10^{-1} $  &    3.807$ \times 10^{-1} $  &   -7.555$ \times 10^{-2} $  &    7.403$ \times 10^{-3} $  \\
$\sigma^+$ &   -11.550 &    1.059  &    4.188$ \times 10^{-1} $  &   -9.351$ \times 10^{-2} $  &    1.242$ \times 10^{-2} $  \\
$\pi$ &   -8.516  &    5.064$ \times 10^{-1} $  &    4.214$ \times 10^{-1} $  &   -8.157$ \times 10^{-2} $  &    9.046$ \times 10^{-3} $  \\
$\pi$ &   -11.470 &    9.645$ \times 10^{-1} $  &    4.605$ \times 10^{-1} $  &   -9.970$ \times 10^{-2} $  &    1.410$ \times 10^{-2} $  \\
$\pi$ &   -8.558  &    5.517$ \times 10^{-1} $  &    4.017$ \times 10^{-1} $  &   -7.866$ \times 10^{-2} $  &    8.149$ \times 10^{-3} $  \\
$\sigma^-$ &   -11.380 &    8.720$ \times 10^{-1} $  &    5.007$ \times 10^{-1} $  &   -1.056$ \times 10^{-1} $  &    1.577$ \times 10^{-2} $  \\
$\sigma^-$ &   -8.474  &    4.600$ \times 10^{-1} $  &    4.415$ \times 10^{-1} $  &   -8.453$ \times 10^{-2} $  &    9.867$ \times 10^{-3} $  \\
$\sigma^-$ &    1.566$ \times 10^{-2} $  &   -4.191$ \times 10^{-3} $  &   -5.948$ \times 10^{-4} $  &    2.399$ \times 10^{-4} $  &    2.909$ \times 10^{-3} $  \\
\enddata
\end{deluxetable}

%%%%%%%%%%%%%%%%%%%%%%%%%%

\begin{deluxetable}{lccccc}
\tablewidth{0pt}
\tablecaption{Component strengths. Tr3}
\label{str_Tr3}
\tablehead{Transition & $d_1$ & $d_2$ & $d_3$ & $d_4$ & Max error }
\startdata
$\sigma^+$ &    0.000  &    0.000  &    0.000  &    0.000  &    0.000  \\
$\sigma^+$ &    1.019$ \times 10^{-1} $  &   -4.220$ \times 10^{-3} $  &   -1.334$ \times 10^{-2} $  &    2.952$ \times 10^{-3} $  &    1.463$ \times 10^{-3} $  \\
$\sigma^+$ &    6.786$ \times 10^{-2} $  &    1.435$ \times 10^{-2} $  &   -1.584$ \times 10^{-2} $  &    2.904$ \times 10^{-3} $  &    1.012$ \times 10^{-3} $  \\
$\pi$ &   -1.019$ \times 10^{-1} $  &    4.220$ \times 10^{-3} $  &    1.334$ \times 10^{-2} $  &   -2.952$ \times 10^{-3} $  &    1.464$ \times 10^{-3} $  \\
$\pi$ &    2.606$ \times 10^{-4} $  &   -3.669$ \times 10^{-2} $  &    1.444$ \times 10^{-2} $  &   -1.789$ \times 10^{-3} $  &    4.060$ \times 10^{-4} $  \\
$\pi$ &    1.019$ \times 10^{-1} $  &   -4.220$ \times 10^{-3} $  &   -1.334$ \times 10^{-2} $  &    2.952$ \times 10^{-3} $  &    1.463$ \times 10^{-3} $  \\
$\sigma^-$ &   -6.812$ \times 10^{-2} $  &    2.232$ \times 10^{-2} $  &    1.415$ \times 10^{-3} $  &   -1.117$ \times 10^{-3} $  &    1.422$ \times 10^{-3} $  \\
$\sigma^-$ &   -1.019$ \times 10^{-1} $  &    4.220$ \times 10^{-3} $  &    1.334$ \times 10^{-2} $  &   -2.952$ \times 10^{-3} $  &    1.464$ \times 10^{-3} $  \\
$\sigma^-$ &    0.000  &    0.000  &    0.000  &    0.000  &    0.000  \\
\enddata
\end{deluxetable}

\begin{figure*}
\epsscale{1.5}
\plotone{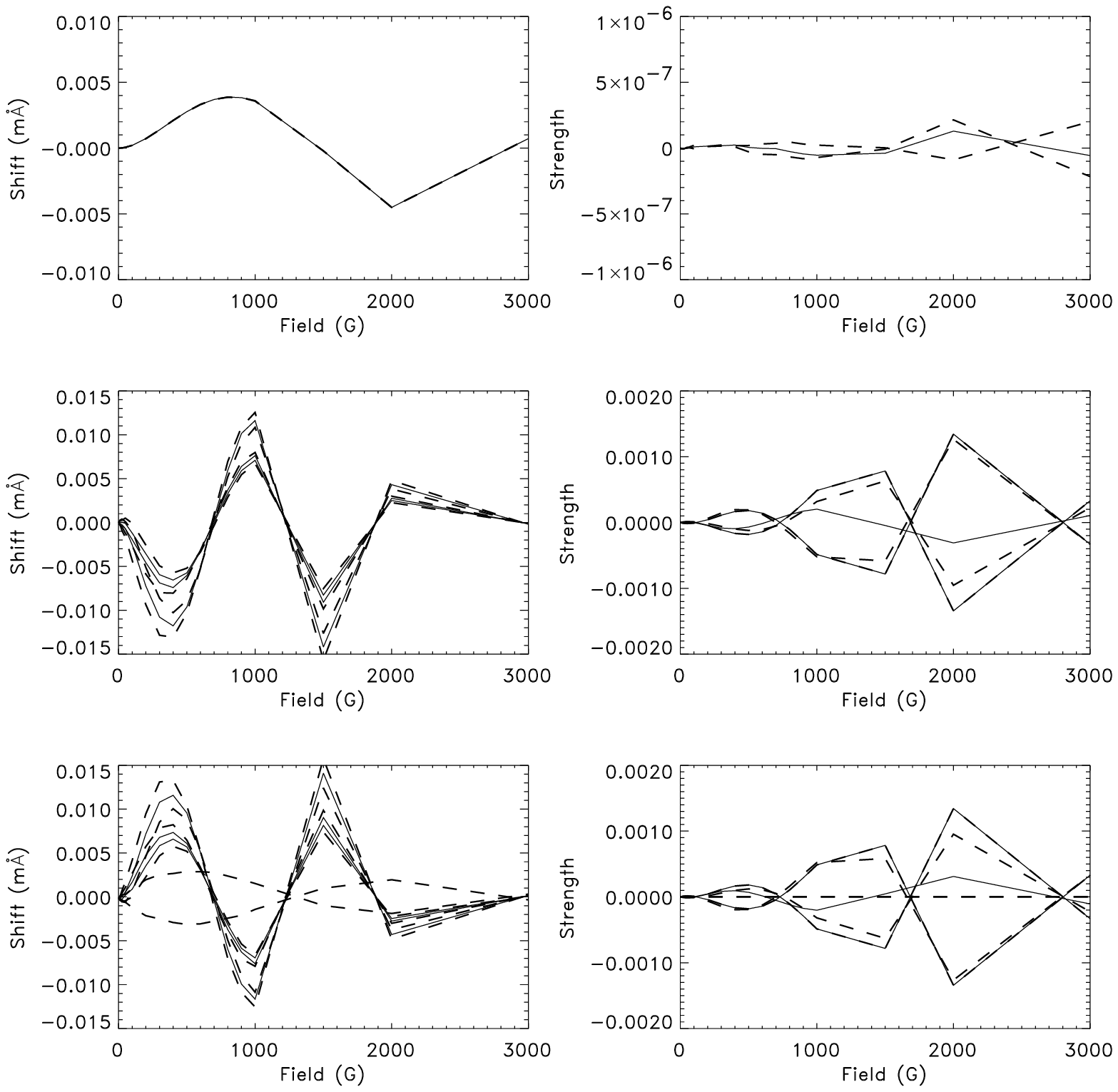}
\caption{Approximation errors. Left panels: Position
  shifts [$(\Delta \lambda_{IPB} - \Delta \lambda_{LZS}) - y_p$]. Right
  panels: Relative strengths [$( f_{IPB} - f_{LZS}) - y_f$]. Top to bottom:
  Tr1, Tr2 and 
  Tr3. In all panels, the solid lines represent $\pi$-components and the
  dashed lines represent $\sigma$-components.
\label{f3}
}
\end{figure*}

\section{Conclusions}
\label{conc}

The polynomial coefficients provided in this work can be used to generalize
existing Zeeman transfer codes to treat the \ion{He}{1} 10830~\AA \,
multiplet in the IPBS regime. As we pointed out in a previous work
(Socas-Navarro et al 2004), neglecting IPBS results in significant errors in
the calculation of its polarization profiles, even for field strengths much
lower than the level-crossing field. Computer codes developed using these
polynomials will be virtually as fast as those using the LZS approximation
but the polarization profiles will be much more accurate. 
% HSN
The actual improvement in CPU time obviously depends on the particular
details of the calculation. In our codes, the use of the polynomials is
approximately a factor 20 faster than the full IPBS implementation.

\acknowledgments
This research has been partly funded by the Ministerio de Educaci\'on y
Ciencia through project AYA2004-05792 and by 
the European Solar Magnetism Network (contract HPRN-CT-2002-00313). 

%\bibliographystyle{../bib/apj}
%\bibliography{../bib/aamnem99,../bib/articulos}

\end{document}